\documentclass[pra,twocolumn]{revtex4}
\usepackage{graphicx}
\usepackage{amsmath}

\begin{document}

\bibliographystyle{unsrt}

\date{\today}

\title{Transmission characteristics of optical nanofibers in metastable xenon}
\author{H.P. Lamsal}
\author{J.D. Franson}
\author{T.B. Pittman}
\affiliation{Physics Department, University of Maryland Baltimore
County, Baltimore, MD 21250}

\begin{abstract}

We study the transmission characteristics of sub-wavelength diameter silica optical nanofibers (ONF's) surrounded with a xenon plasma produced by a low-pressure inductive RF discharge. In contrast to related experiments using rubidium vapor, we find essentially no degradation of optical transmission through the ONF's as a function of time. We also observe a pronounced ONF transmission modulation effect that depends on the conditions of the xenon plasma.
\end{abstract}

\pacs{XYZ}

\maketitle

The use of optical nanofibers (ONF's) in warm atomic vapors has recently been identified as a robust platform for ultralow-power nonlinear optics \cite{spillane2008,nieddu2016,solano2017}.  In these systems, the ONF can be viewed as a sub-wavelength diameter cylindrical silica waveguide that supports the propagation of a tightly confined evanescent mode through the surrounding atomic vapor \cite{snyderlovebook,tong2004}. The diameter of the ONF can be chosen to optimize the evanescent mode profile for various nonlinear interactions \cite{you2008}, and the choice of atoms can dictate the operating wavelength for practical ultralow-power nonlinear optics devices.

To date, nearly all experimental work in this area has used rubidium as the atomic vapor (see, for example, \cite{spillane2008,hendrickson20010,salit2011,watkins2013,jones2015}). This opens the question of whether other atomic species may be compatible, or even beneficial, in the  basic ``ONF in warm atomic vapor'' platform.  A recent experiment explored the use of metastable xenon (Xe*) as a promising alternative to Rb \cite{pittman2013}.  Rubidium has a tendency to adsorb to silica surfaces \cite{ma2009}, and the accumulation of Rb on the ONF surface can lead to scattering of the evanescent mode and a consequent loss of overall transmission through the ONF. This detrimental loss mechanism can be quite large, and can occur on timescales of only a few minutes under typical experimental conditions \cite{lai2013}. The primary motivation for using Xe (an inert nobel gas) rather than Rb (a reactive alkali) was to avoid this type of rapid transmission loss  \cite{pittman2013}. 

Although that preliminary experiment did demonstrate ultralow-power nonlinearities using ONF's in Xe*, it was not able to investigate the platform's long-term transmission characteristics due to technical limitations associated with the DC discharge system used to promote the Xe atoms to the metastable state.   Consequently, a study of the transmission characteristics of ONF's in Xe* is currently of great interest.  

Here we describe such a study using Xe* produced by a low-pressure inductive RF discharge system. In contrast to the case of Rb, we find that  ONF's in Xe* suffer essentially no transmission degradation as a function of time. We also observe an ONF transmission modulation effect that can be controlled by the RF discharge system. 

\begin{figure}[b]
\includegraphics[width=3.4in]{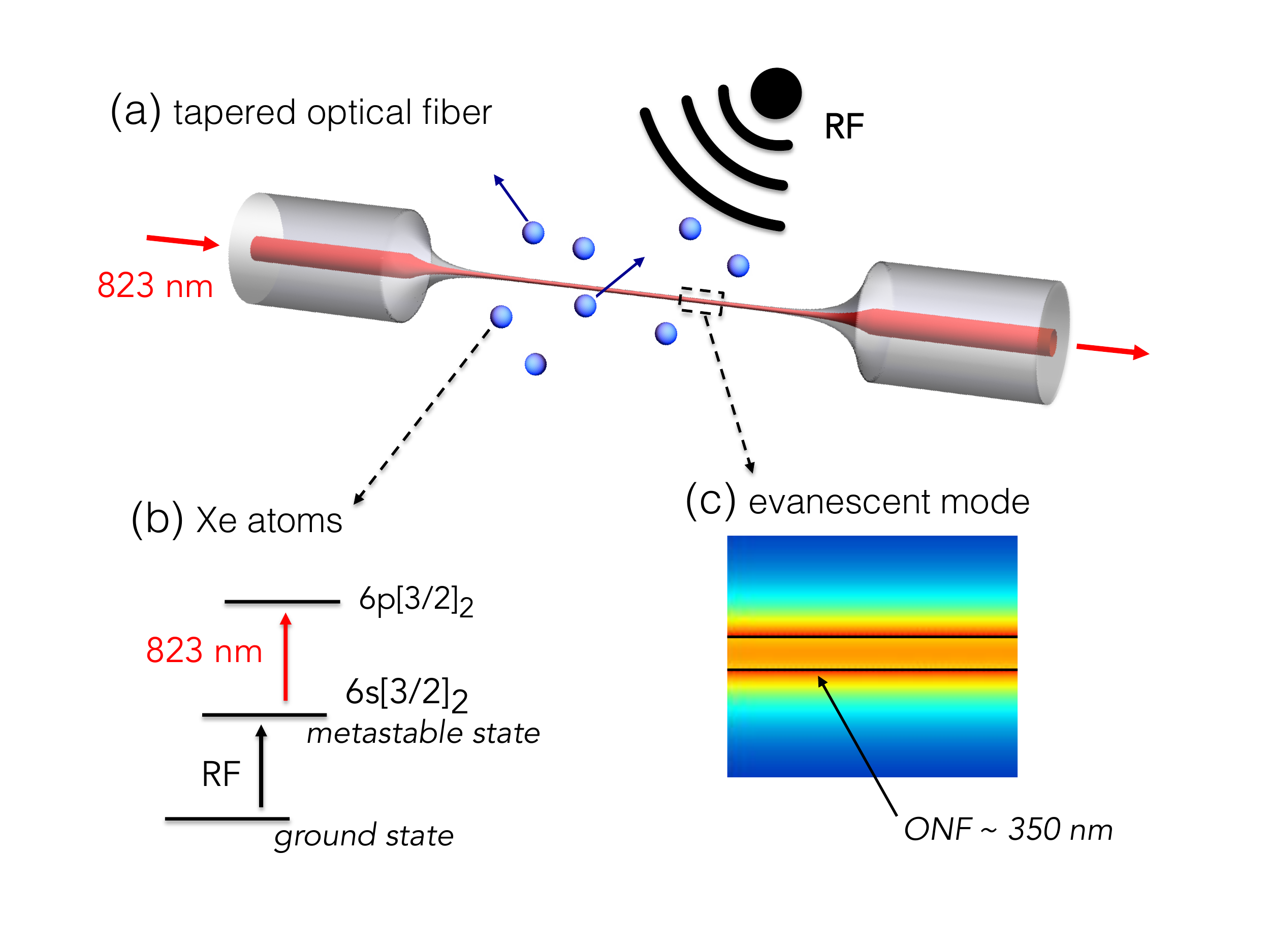}
\caption{(Color online) Overview of the basic ONF-Xe* platform. (a) an ONF formed in the waist of a tapered optical fiber (TOF) is surrounded by a gas of Xe atoms. The untapered ends of the TOF provide convenient fiber-coupled input and output ports for the measurements of interest. (b) a strong RF field is applied to promote the Xe atoms to the metastable state, which serves as an ``effective ground state'' for optical absorption experiments at 823 nm. (c) calculation of the evanescent mode of a 350 nm diameter ONF (denoted by the black lines).  A signifiant portion of the mode propogates outside of the ONF, where it can interact with the surrounding Xe* atoms \protect\cite{snyderlovebook}.}
\label{figure:fig1}
\end{figure}

\begin{figure}[t]
\includegraphics[width=3.4in]{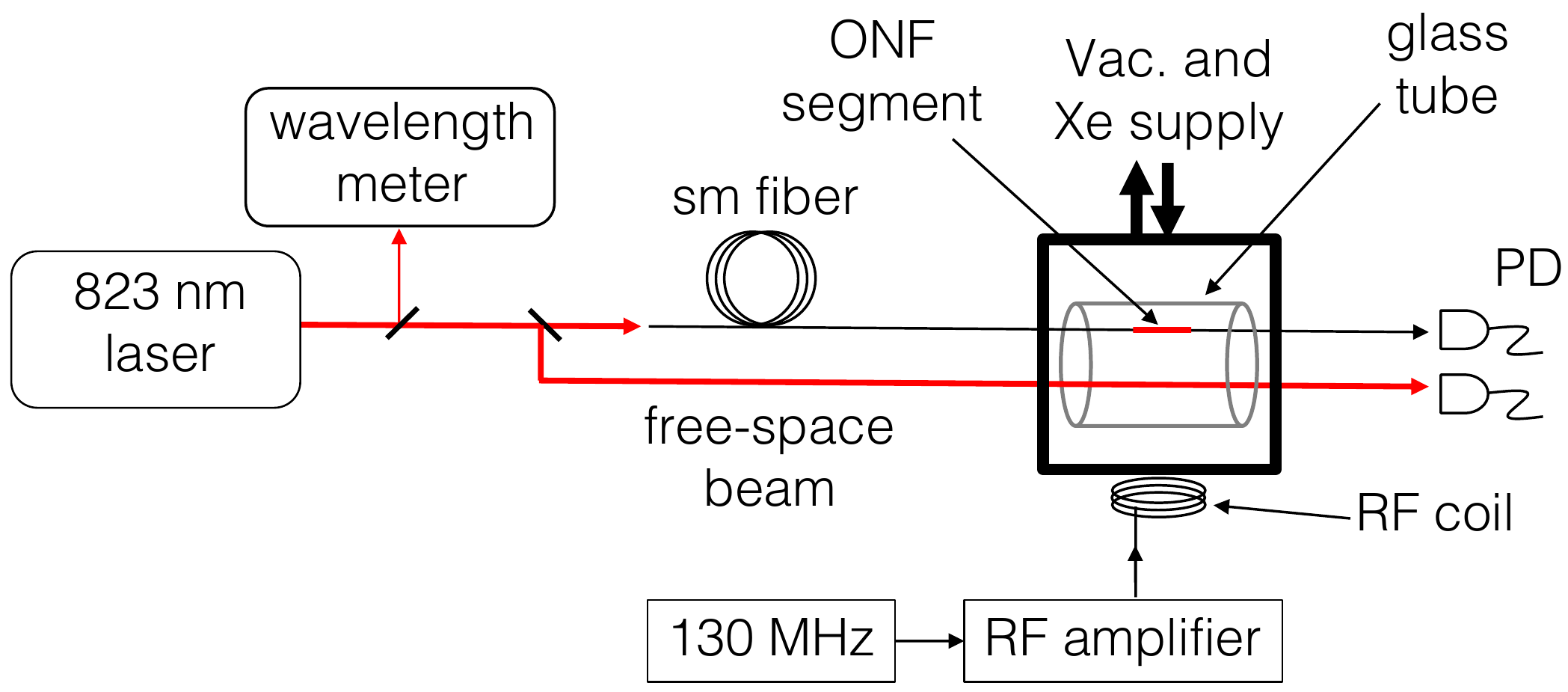}
\caption{(Color online) Overview of the experimental arrangement. The TOF (with ONF region) is installed in a vacuum system that is back-filled with controlled pressures of Xe.  An external RF discharge coil (mounted on a glass window) is used to create a Xe plasma inside the chamber. The output of a narrowband tunable diode laser (at 823 nm) is split to provide the input to the TOF (standard single-mode fiber), as well as an auxiliary free-space beam. Photodiodes (PD) are used to simultaneously measure both the ONF and free-space transmission signals. A loose-fitting glass tube is used to protect the ONF from contaminants released from the vacuum chamber walls during RF discharges. }
\label{fig:fig2}
\end{figure}

An overview of the system under study in shown in Figure \ref{figure:fig1}. ONF's with a diameter of  $\sim$350 nm and a length of $\sim$5 mm are formed in the waist of a tapered optical fiber (TOF) using the flame-brush technique \cite{birks1992}. The TOF is installed in a vacuum system that is back-filled with Xe gas at pressures around 30 mTorr.  A strong RF field is then applied to produce a glow discharge plasma, which promotes a fraction of the Xe atoms to the $6s[3/2]_{2}$ metastable state \cite{pittman2013}. This metastable state serves as a long-lived  ``effective ground state''  for various optical absorption spectroscopy experiments using the $6s[3/2]_{2} \rightarrow 6p[3/2]_{2}$ transition at 823 nm \cite{walhout1994}. Broadly speaking, the absorption signal (measured at the output of the TOF) provides a measure of the atom-field interactions that occur when Xe* atoms pass through the tightly confined evanescent mode in close proximity to the ONF surface.

Figure \ref{fig:fig2} shows a block diagram of the experimental arrangement used to characterize ONF transmission in this system.  The TOF is installed inside an ultra-high vacuum chamber formed by a standard 4.5'' conflat (CF) cube. The fiber leads (far from the TOF region) are fed through the vacuum chamber using CF Swagelok feedthrough flanges with teflon ferrules \cite{abraham1998}. A  2'' diameter glass tube is placed around the TOF region inside the vacuum chamber. This loose-fitting glass tube allows Xe gas to easily flow into the TOF region, but protects the ONF from any contaminants released from the chamber walls during the RF discharge process.   

Optical viewports (windows) are used on the chamber to enable an auxiliary free-space beam to propagate through the glass tube next to the ONF region. The simultaneous Xe* spectroscopic absorption signals from this auxiliary free-space beam and the ONF evanescent guided-mode are used to monitor the metastable state density in the chamber and characterize the ONF transmission as a function of time.

The RF discharge system is based on a standard LC tank circuit formed by a coil and capacitors with a resonance frequency of  $\sim$130 MHz. The coil is placed on the outside of a viewport located on the side of the vacuum chamber.  We drive this system with a variable-power 130 MHz signal passing through a 25 W broadband linear RF amplifier.  The RF field produces a glow discharge plasma inside the glass tube and the rest of the vacuum chamber.

\begin{figure}[b]
\includegraphics[width=3.43in]{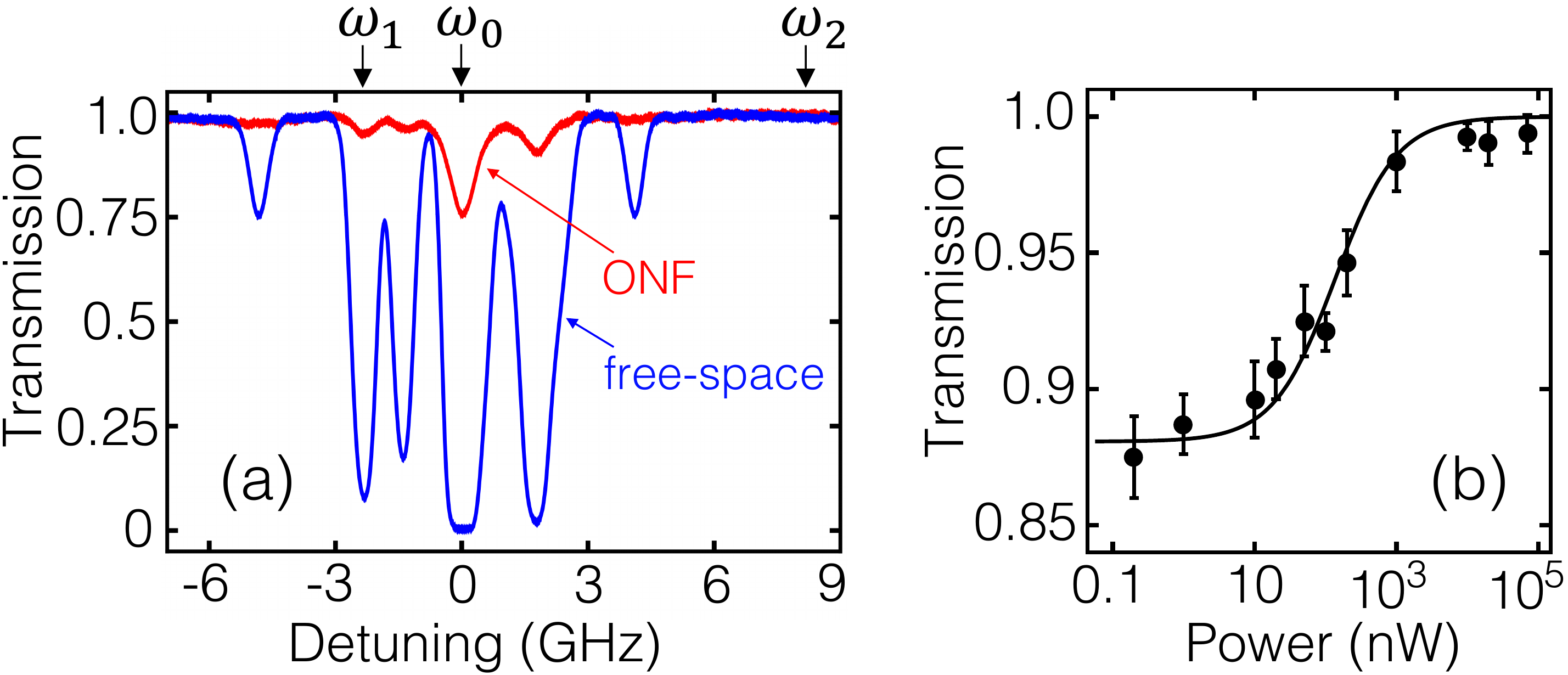}
\caption{(Color online) (a) simultaneously measured absorption spectra for the auxiliary free-space beam (blue trace), and the ONF evanescent mode (red trace). The detunings are relative to a central frequency of $\omega_{0}$ = 364.095 THz. The arrows denote three specific frequencies that will be used to characterize transmission in the system. Panel (b) shows the ability to saturate absorption (at $\omega_{0}$) at remarkably low power levels ($\sim$10 nW) using the ONF-Xe* platform  \protect\cite{spillane2008,pittman2013}. The solid line is a fit to a simple saturation model \protect\cite{jones2014}. }
\label{fig:fig3}
\end{figure}

The output of a tunable diode laser centered at 823 nm (Toptica DL-Pro, linewidth $\sim$300 kHz) is split to provide both the free-space and ONF signals in the system.  Figure \ref{fig:fig3} shows examples of the various absorption signals measured under typical operating conditions with the RF discharge applied.  The blue trace in panel (a) shows the absorption spectra of the auxiliary free-space beam as the laser frequency is swept across the relevant Xe* resonance at 823 nm. The 6 absorption resonances are due to various hyperfine splittings of the $6s[3/2]_{2} \rightarrow 6p[3/2]_{2}$ transition for the 9 isotopes of natural Xe \cite{xia2010}.   

The red trace in panel (a) shows the corresponding signal that was simultaneously obtained using the ONF.  Here, the observation of the same 6 absorption resonances is an indication of the interaction of the ONF evanescent mode with the surrounding Xe* atoms.  The smaller absorption features in the ONF case are primarily due to the shorter effective interaction length of the ONF ($\sim$5 mm) compared to that of the free-space beam ($\sim$100 mm). The existence of small transit-time broadening effects due to the tightly confined evanescent mode area of the ONF also play a role in the spectral lineshapes \cite{spillane2008,jones2014}.   The small arrows in Figure \ref{fig:fig3}(a) mark two specific ``on resonance'' frequencies, $\omega_{0}$ and $\omega_{1}$, that will be used to monitor the Xe* density in the system, and the ``off resonance'' frequency  $\omega_{2}$ that will be used to monitor the overall transmission through the ONF.

Figure \ref{fig:fig3}(b) shows the ability to saturate the $6s[3/2]_{2} \rightarrow 6p[3/2]_{2}$ transition in Xe at remarkably low power levels ($\sim$10 nW) using an ONF.  Here the laser is tuned to the on-resonance frequency $\omega_{0}$, and the power is gradually increased to reveal a nonlinear response in the atomic absorption. This basic saturation effect provides a benchmark of the strength of the optical nonlinearity in the system and confirms the utility of the ONF-Xe* platform for ultralow-power nonlinear optics \cite{pittman2013}. It is worth noting that the saturation power of $\sim$10 nW observed in this ONF-Xe* system is comparable to that observed in typical ONF-Rb systems \cite{spillane2008,hendrickson20010,jones2014}.

\begin{figure}[t]
\includegraphics[width=3.45in]{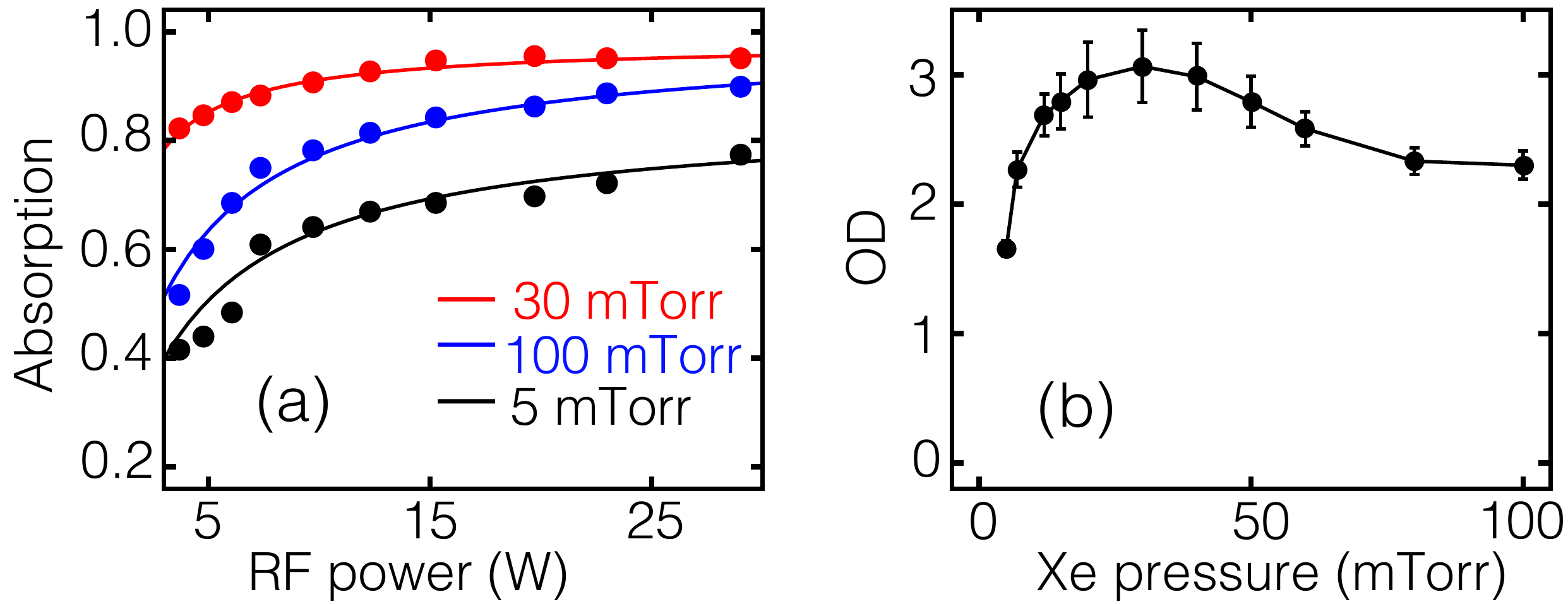}
\caption{(Color online) Dependence of overall Xe* density on RF discharge power and neutral Xe pressure. (a) free-space beam absorption (at  $\omega_{1}$) as a function of the RF power applied to the discharge coil, for 3 different neutral Xe pressures. The solid lines are fits to a simple saturation model. (b) measured optical depth (OD) for the free-space beam (at $\omega_{1}$) as a function of Xe pressure, using a nominal RF power of 20 W. We obtain the maximum Xe* density using a Xe pressure of $\sim$30 mTorr in our system.}
\label{fig:fig4}
\end{figure}

In order to study the long-term transmission characteristics of ONF's in Xe*, we first optimize our RF discharge system to promote the maximum possible fraction of Xe atoms to the metastable state.  Figure \ref{fig:fig4} provides a measure of the Xe* density as function of RF power for several neutral Xe pressures.  In our system, we observe only marginal increases in the Xe* density at RF powers greater than $\sim$15 W (Figure \ref{fig:fig4}(a)), and an optimal neutral Xe pressure of $\sim$30 mTorr  (Figure \ref{fig:fig4}(b)). The self-limiting effects seen in Figure \ref{fig:fig4} are fairly typical in these types of RF discharge systems, with the maximum achievable Xe* density primarily being limited by collisions and other competing processes in the plasma \cite{xia2010}.

We obtain estimates of our  Xe* density by fitting free-space absorption spectra lineshapes using the models of reference \cite{fitting}, while taking into account the natural abundances of the various Xe isotopes and branching ratios of the hyperfine transitions \cite{branchingratios}. Using these techniques, we estimate the maximum Xe* density to be $\sim$10$^{11}$ cm$^{-3}$ in our particular system.

\begin{figure}[b]
\includegraphics[width=3.35in]{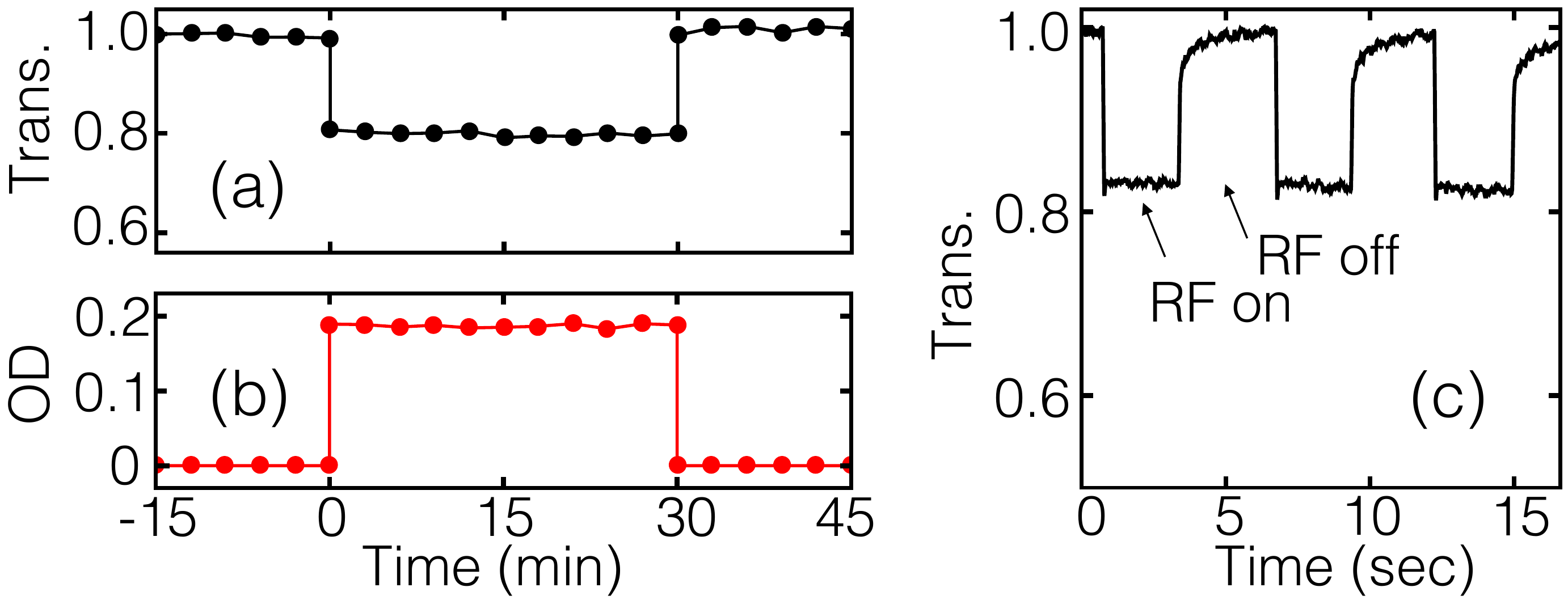}
\caption{(Color online) Typical results of ONF transmission vs. time measurements using our maximum Xe* density. (a) normalized overall TOF transmission (at $\omega_{2}$). Here, the RF discharge is turned on at $t=0$ min. and off at $t=30$ min. (b) ONF OD (at $\omega_{1}$) indicates a constant Xe* density during the 30 minute run. (c) TOF transmission (at $\omega_{2}$) on an arbitrarily-chosen timescale highlighting a pronounced modulation effect that can be controlled by simply turning the RF discharge on and off.}
\label{fig:fig5}
\end{figure}

Figure \ref{fig:fig5}(a) shows a typical measurement of ONF transmission vs. time in the system using our maximum Xe* density.  ONF spectra (similar to that shown in Figure \ref{fig:fig3}(a)) are recorded at 3 minute intervals, with absorption values at $\omega_{2}$ used to measure the overall off-resonant ONF transmission, and absorption values at $\omega_{0}$ used to characterize the ONF optical depth (OD) during the run (Figure \ref{fig:fig5}(b)).  The overall transmission is normalized to unity at the beginning of a run by the maximum system transmission of $\sim$70\% measured after preliminary TOF fabrication and installation in the vacuum chamber \cite{99percent}.  The data in the central region of Figure \ref{fig:fig5} represent the results of a typical 30 minute run with constant Xe* density, during which time the ONF suffered essentially no loss of transmission. This long-lived lossless transmission characteristic is the key result of the study.

An auxiliary result can be seen by the temporary drop and complete recovery of transmission at the beginning and end of the 30 minute run. This transmission modulation, which is controlled by simply turning the RF discharge on and off, is highlighted in Figure \ref{fig:fig5}(c). The origin of this effect is most likely due to a combination of refractive index changes and ONF heating effects.  Roughly speaking, high transmission through the tapered regions of a TOF is governed by an adiabaticity requirement that critically depends on the contrast between the core and cladding refractive indices \cite{solano2017}. The creation of a Xe plasma provides a relatively large change in the effective refractive index for the ``cladding'' of the TOF \cite{plasmaindex}, while discharge-based heating of the ONF region may also result in small changes to the core index.  We found that the modulation depth ($\sim$20\% in Fig. \ref{fig:fig5}) was always constant during a 30 minute run, but varied between values of $\sim$15 - 25\% from run to run due to the establishment of different discharge plasma conditions.  We were unable to characterize the maximum possible modulation frequency above $\sim$100 Hz due to the slow response time of the detectors used in our experiments, and difficulties in rapidly switching the RF discharge system.

We typically limited the duration of our transmission-study runs to 30 minutes due to a gradual drop in Xe* density in our closed system for longer times.  At the end of these runs the system could be quickly pumped out and re-backfilled with another $\sim$30 mTorr of Xe, enabling a rapid series of 30 minute runs using the same ONF.  We used this technique to study single ONF's under cumulative Xe plasma exposure times of several hours and also observed no transmission loss for these longer term studies.
 
These results indicate the ability to perform long-term ultralow-power nonlinear optics experiments using ONF's in Xe* without any degradation in transmission. The observed transmission characteristics support the intuitive idea that Xe, an inert noble gas, does not accumulate on the surface of the ONF like Rb, a reactive alkali metal. In addition, the results suggest that significant plasma-etching of the silica itself (resulting in a ``rough ONF surface'' and loss of transmission) does not occur under these  mild conditions of a low-pressure isotropic RF Xe discharge  \cite{cardinaud2000}.

Although the Xe* densities of $\sim$10$^{11}$ cm$^{-3}$  used here are lower than the typical densities of $\sim$10$^{12}$ cm$^{-3}$ used in related Rb experiments, the complete lack of transmission degradation over very long time-scales suggests the ability to use ONF's in higher Xe* densities as well.  We note that higher Xe* densities may be possible using optical pumping techniques \cite{hickman2016}, and these techniques may also be relevant for the ONF-Xe* platform \cite{lamsal2019}. In addition, we note that we have previously observed the onset of ONF transmission degradation using Rb densities even lower than $\sim$10$^{11}$ cm$^{-3}$ \cite{lai2013}.
 
Consequently, for practical applications requiring the highest OD's, but where high-transmission is not critical, the use of Rb and various LIAD- and/or heating-based techniques to mitigate Rb accumulation may be optimal \cite{spillane2008,hendrickson2009,lai2013}. In contrast, for applications requiring the highest possible long-term ONF transmission, but where the highest OD's are not critical, the use of Xe* may be beneficial. This latter class of applications may include, for example, experiments using various types of nanofiber-based optical resonators \cite{wuttke2012,kato2015,edwards2014,jones2016}.

In summary, we have studied the transmission characteristics of ONF's in Xe* produced by an RF discharge, and observed no degradation in transmission as a function of time.  In addition, we have identified a pronounced modulation effect that can be controlled by the RF discharge itself. These results help extend the first observations of atom-field interactions using ONF's in Xe* \cite{pittman2013} towards a practical long-term and robust system for ultralow-power nonlinear optics.

{\bf Acknowledegments:} This work was supported in part by the National Science Foundation under grant No. 1402708 and the Office of Naval Research under grant N00014-15-1-2229.


\end{document}